\documentclass[letterpaper]{article}
\usepackage{aaai2026} % DO NOT CHANGE THIS
\usepackage{times} % DO NOT CHANGE THIS
\usepackage{helvet} % DO NOT CHANGE THIS
\usepackage{courier} % DO NOT CHANGE THIS
\usepackage[hyphens]{url} % DO NOT CHANGE THIS
\usepackage{graphicx} % DO NOT CHANGE THIS
\usepackage{graphicx} % DO NOT CHANGE THIS
\usepackage{natbib} % DO NOT CHANGE THIS
\usepackage{caption} % DO NOT CHANGE THIS
\usepackage{acronym}
\usepackage[inline]{enumitem}
\usepackage{csquotes}

\title{Bathtubs, Boundaries, and Sandboxes: AI Regulatory Learning under Legal Uncertainty}
\author {
    Tom Deckenbrunnen\textsuperscript{\rm 1,2},
    Alessio Buscemi\textsuperscript{\rm 1},
    Marco Almada\textsuperscript{\rm 2},
    Alfredo Capozucca\textsuperscript{\rm 2},
    German Castignani\textsuperscript{\rm 1}
}
\affiliations {
    \textsuperscript{\rm 1}Luxembourg Institute of Science and Technology\\
    \textsuperscript{\rm 2}University of Luxembourg\\
    \{tom.deckenbrunnen, alessio.buscemi, german.castignani\}@list.lu, marco.almada@cunef.edu, alfredo.capozucca@uni.lu
}
\begin{document}

\maketitle

\begin{abstract}
Effective regulation of AI is a defining policy challenge, driven by its integration into all aspects of society.
To remain responsive to the rapid development and emergent properties of AI systems, policymakers across the globe rely on high-level principles and abstract legal requirements.
While this flexibility supports future-proofing human-centred regulations and aligning them with socio-ethical values, it causes legal uncertainty downstream as developers, companies, and auditors struggle with translating these abstract requirements into verifiable technical properties.
Using the AI Act as an example, this paper draws on Coleman’s bathtub to analyse the regulatory learning space in AI governance.
It argues that legal uncertainty cannot be fully reduced ex ante and that, within reasonable bounds, it is also necessary for regulatory learning because it creates the space in which boundary negotiation over socio-technical meaning can occur.
Building on this analysis, the paper shows how boundary objects and boundary negotiating artifacts help explain the translation of legal requirements into operational practice.
By examining technical sandbox frameworks, it further identifies concrete properties that technical infrastructures must possess to function effectively as boundary negotiation artifacts in AI assessment.
The paper concludes that legal certainty remains the long-term aim, but that premature closure of regulatory instruments risks undermining the learning processes needed for adaptive governance.
\end{abstract}


\section{Introduction}
\label{sec:introduction}

The regulation of AI has become a defining policy challenge of the digital era.
As AI systems permeate economic, social, and political domains, governments worldwide face the difficult task of balancing innovation with protection, flexibility with accountability, and global competitiveness with domestic values.
Despite a shared sense of urgency, national and regional approaches to AI governance diverge, reflecting distinct legal traditions and philosophies of regulation.
The \ac{EU} stands out for its historically anticipatory and top-down regulatory tradition, exemplified by the \ac{AI Act}~\cite{EUAIAct2024}.
Unlike jurisdictions such as the United States, which rely primarily on sectoral rules, voluntary standards, and ex-post liability~\cite{us_ai_action_plan_2025}, or China, where governance is framed through industrial policy and targeted regulatory interventions in prominent domains~\cite{chenStateSocietyMarket2025}, the EU has opted for a comprehensive, fundamental rights-based, and risk-oriented framework~\cite{almadaEUAIAct2025}.
The \ac{AI Act} imposes horizontal \textit{ex-ante} obligations, especially for high-risk systems, spanning \textit{inter alia} risk management, data governance, and technical documentation.

Despite diverging approaches, these legislations share a common concern in the translation of abstract legal and regulatory requirements into verifiable technical practices and tests, causing widespread uncertainty in operationalising regulations \cite{regulatinguncertainty2025}.
This regulatory uncertainty is increasingly recognised as a practical compliance concern for developers, companies, and auditors alike~\cite{lewis2025regulatorylearning,schiffEmergenceArtificialIntelligence2024, schiffStrategiesHarmonizingFragmented2025}.
A growing body of studies and surveys indicates that this uncertainty generates hesitation in investment, uneven compliance readiness, and fragmented approaches to risk management~\cite{deloitte2024ai,dlapiper2025pause,draghi2024competitiveness,arnal2024implementation,regulatinguncertainty2025}.
At the same time, the interpretative effort to operationalise legally uncertain requirements presents a set of opportunities.

The \ac{AI Act} is a distinctive case among legislations and regulations.
As part of the \ac{NLF}, it responds to the limitations of purely legislative governance by introducing a suite of formal instruments and mechanisms designed to foster regulatory learning and innovation support, signalling a gradual move toward bottom-up governance \cite{crumBrusselsEffectExperimentalism2025}.
These include Codes of Practice (Art.~56), \acp{AIRS} (Arts.~57–58), controlled data access and real-world testing environments (Arts.~59–61), and flexibility measures for specific operators such as \acp{SME} (Art.~63).
The inclusion of such provisions embodies a shift in mindset in which regulators are expected not only to enforce compliance but also to listen to practitioners, learn from experimentation, and adapt rules accordingly \cite{crumBrusselsEffectExperimentalism2025, ruschemeierExperimentalRegulationAI2025}.
Yet, the widely disputed and recently adopted Digital Omnibus contains a multitude of provisions for further specifying amendments of \acp{AIRS}, showing that the uncertainty extends to the legal framework itself~\cite{DigitalOmnibus2026}.

Beyond the adaptive legal framework and its attempts to codify human values, AI systems are deeply embedded in socio-technical contexts, both through their training data and the environments they are deployed in.
This embeddedness means the interpretation of the legal requirements is itself a socio-technical process shaped by the involvement of different stakeholder groups.
Under this premise, this paper contributes to AI ethics and society research by developing a socio-technical account of regulatory learning using the EU \ac{AI Act} as a concrete example.
It argues that legal uncertainty cannot be fully reduced ex ante and that, within reasonable bounds, it is also necessary for regulatory learning, because it creates the space in which abstract requirements must be negotiated and translated into operational practice.
To analyse this process, the paper draws from sociology, via Coleman's bathtub and boundary objects, and shows how the \ac{AI Act}, through legislative pressure and legal uncertainty, creates a space for negotiation and contestation of abstract legal requirements across stakeholder communities.
It further argues that the translation of legal into technical requirements can be more deeply understood through \acp{BNA}.
Finally, by focusing on \acp{AIRS} and related technical frameworks, it shows how this multi-stakeholder boundary negotiation may be operationalised in practice and identifies concrete properties that technical infrastructures must possess to function effectively as \acp{BNA}.

\section{Policy Learning and Coleman's Bathtub}
\label{sec:policy-learning}

The AI Act, adopting a general definition of AI system as any machine-based system with varying levels of autonomy that may exhibit adaptiveness after deployment and that infers how to generate outputs from its inputs (Article 3), casts a large net to ensure wide coverage of the regulation \cite{EuropeanCommission2025Guidelines}.
Within this broad scope, modern generative AI systems pose a particular challenge due to their emergent, i.e., not explicitly programmed, behaviour arising from the complex interplay of their components, causing parallels to be drawn to established complex system sciences \cite{holtzmanGenerativeModelsComplex2025}.
Reflecting that reality, these considerations are extended from the development and analysis of AI systems to their governance \cite{koltLessonsComplexSystems2025a}.
To retain agility in its fundamental rights, risk-based approach in light of such unpredictable developments, the AI Act is integrated into the \ac{NLF} \cite{vealeDemystifyingDraftEU2021}.
It sets horizontal essential requirements and delegates technical specifications to harmonised standards and implementing and delegated acts for vertical governance \cite{leydenStandardsEUAI2025}.
However, a critical gap currently exists as the vertical, sector-specific standards required for practical implementation, to be developed by bodies such as \ac{CEN} and \ac{CENELEC}, have yet to be adopted by the \ac{EU} Commission \cite{leydenStandardsEUAI2025, UpdateCENCENELECs}.

Despite this current lack of standards, the AI Act, with provisions enabling regulatory learning such as the mandatory establishment of regulatory sandboxes in each Member State, reporting mandates, and regular reviews of the legislation, is intended to be a framework of future-responsive, adaptive regulation integrating both top-down enforcement and bottom-up learning \cite{schrepelAdaptiveRegulation2025a,ebersTrulyRiskbasedRegulation2025}.
The governance structure of the \ac{EU}, especially in its AI regulatory space, forms a complex multi-level network of actors and their interactions~\cite{novelliRobustGovernanceAI2025}.
An analysis of regulatory learning, how learning signals are produced and cause change in regulation, can therefore be difficult to track across individual actor relationships, requiring a certain degree of simplifying abstraction.
In the extant literature on regulatory theory and policy learning, this bi-directional information flow is described as taking place between two different levels, namely the \textit{macro} level of societal learning processes and the \textit{micro} level of individual cognitive processes~\cite{zakiConceptualisingOrganisationalPolicy2025, dunlopLearningBathtubMicro2017}.
At the risk of introducing a competing frame from the \ac{EU}'s actual governance levels, the application of the established instruments decouples the analysis of regulatory learning mechanisms from the governance levels and supports its transferability to other political regimes.

One particularly popular analytical model, drawn from social sciences and capturing the relationship within and between these levels, is Coleman's bathtub~\cite{colemanSocialTheorySocial1986}.
In a policy learning context, it describes the top-down enforcement as macro-micro interaction and the bottom-up feedback by way of micro-macro interactions.
The macro-micro influence applies exogenous pressure, through enforced regulation from the macro level, on individuals at the micro level, forcing them to adapt and undergo their learning cycle.
It is through this process at the micro level that the evidence and feedback to the macro level is generated.

At the same time, the individual micro-level learning process is unable to produce a stimulus large enough to incite learning at the macro level.
Through micro-micro transitions, in which groups of individuals interact, learning becomes a collective process, providing the signal to prompt potential learning at the macro level.
Overall, the model captures the continuous feedback loop between the macro and micro levels by way of the macro-micro, micro-micro, and micro-macro interactions.
It provides an idealised model of how both levels influence each other's learning processes, with the micro level providing the foundations for evidence generation for learning at the macro level.
However, some models for policy and regulatory learning do not neatly fit Coleman's bathtub, due to a missing element mediating between the macro and micro levels, micro-micro transitions not fully capturing the aggregation of learning evidence between levels \cite{dunlopLearningBathtubMicro2017}, or the need for explicit presentation of the intermediate level in organisational policy learning \cite{zakiConceptualisingOrganisationalPolicy2025}.
Publications aiming to supplement the analysis of learning dynamics often introduce or specify the \textit{meso} level to capture these organisational processes that mediate interactions between micro-scale and macro-scale learning \cite{welchConceptualizingMesoLevelOrganizations2025a, jeppersonMultipleLevelsAnalysis2011, moysonPolicyLearningPolicy2017}.
The meso level consists precisely of this aggregation layer at which organisations contribute to and participate in collective learning, ultimately feeding the macro level.

We summarise here the three levels and their roles in the learning cycle:
\begin{itemize}
    \item \textbf{macro:} defines the regulatory frame and applies top-down pressure on the micro level
    \item \textbf{meso:} translates, mediates, and aggregates between the macro and micro levels
    \item \textbf{micro:} absorbs the pressure, adjusts its practices, and produces learning signals
\end{itemize}

\section{Regulatory Learning in the AI Act's Governance Structure}
\label{sec:ai-act-bathtub}

\subsection{Actors and Mechanisms}

The AI Act, through Articles 64 to 70, defines a multi-actor structure aiming at a distributed governance model through the decentralisation of tasks.
It attributes a diverse set of roles in enforcement and reporting to the defined actors, giving rise to complex interactions between them, marking the AI Act as a distinctly collaborative governance framework \cite{cancela-outedaEUsAIAct2024}.
\citeauthor{novelliRobustGovernanceAI2025} have provided a hierarchical analysis of the governance structure, decomposing the actors and their interplay into corporate, national, and supranational levels~\cite{novelliRobustGovernanceAI2025}.
Their institutional analysis primarily focused on aspects of agency and interoperability in the distributed governance model of the \ac{AI Act}.
Recommendations for increasing the robustness of its governance model included the institutional design of the AI Office, i.e.\, its responsibilities and scope, and a central coordination hub for interoperability.
Here, we depart from a structural and institutional analysis and present a simplified account of the regulatory learning space via the bi-directional role the \ac{AI Act}'s actors inhabit within that space.
This is particularly useful in analysing those actors assigned with multiple roles, effectively becoming mediators between the macro and the micro levels, resulting in messy, complex interactions.

This observation can be illustrated by considering the role of the AI Office under Article 64.
The AI Office is situated within the European Commission, under DG CONNECT, and does not constitute a legally independent body.
This institutional placement primarily reflects considerations of coordination and policy coherence.
Simultaneously, it highlights the analytical challenge of distinguishing between governance structures designed for oversight and coordination, and mechanisms explicitly oriented toward the systematic aggregation and circulation of implementation-level evidence within a regulatory learning framework.
Nevertheless, the AI Office is entrusted with a range of responsibilities, including the enforcement of the regulation with respect to \ac{GPAI} models, support for coordination of implementation across Member States, and participation in standardisation processes. Taken together, these tasks suggest a degree of operational discretion in practice, without formal institutional independence.
From an analytical perspective, this positioning can be understood as conferring a form of limited and context-dependent operational autonomy, whose scope and implications for regulatory learning remain to be further clarified in the literature \cite{novelliRobustGovernanceAI2025, cancela-outedaEUsAIAct2024, grauxInterplayAIAct2025}.
This disconnect between the AI Office's legal and operational autonomy makes it a \textit{quasi-agency}, marking an example of agencification, an ongoing trend within the \ac{EU} governance landscape \cite{verhoestAgencificationEurope2018, korverEUAgencies2018,novelliRobustGovernanceAI2025}.

A useful heuristic to apply in positing the bi-directional flow is based on a functional rationale, drawing from the top-down enforcement pipeline with the legislation at its origin.
With the AI Act taking a horizontal approach to regulation, it leaves the instantiation and operationalisation of technical requirements to harmonised standards \cite{gornetEuropeanApproachRegulating2024}, delegated acts or implementing acts \cite{EUComDraftAIRS2025}.
These technical requirements then, in turn, need to be operationalised via specific assessment practices and tests.
As such, the legal requirements from the AI Act are specified at three different levels of abstraction: 
the legislative level, in which the lawmakers establish legal obligations that are binding but cast in general terms to cover a wide range of contexts;
the regulatory level, in which binding instruments (such as implementing acts adopted by the EU Commission) or non-binding texts (such as harmonised technical standards) detail the implications of the legislative commands with regard to specific technologies and contexts;
and the technical level, in which the regulatory commands are translated into software requirements to be implemented into an AI system and technical assessments to be performed on said system.
Learning can take place at each of these three levels of abstraction, giving rise to adaptation in technical practices, applicable standards, and the AI Act itself.

\begin{figure*}[!htb]
    \centering
    \includegraphics[width=1\linewidth]{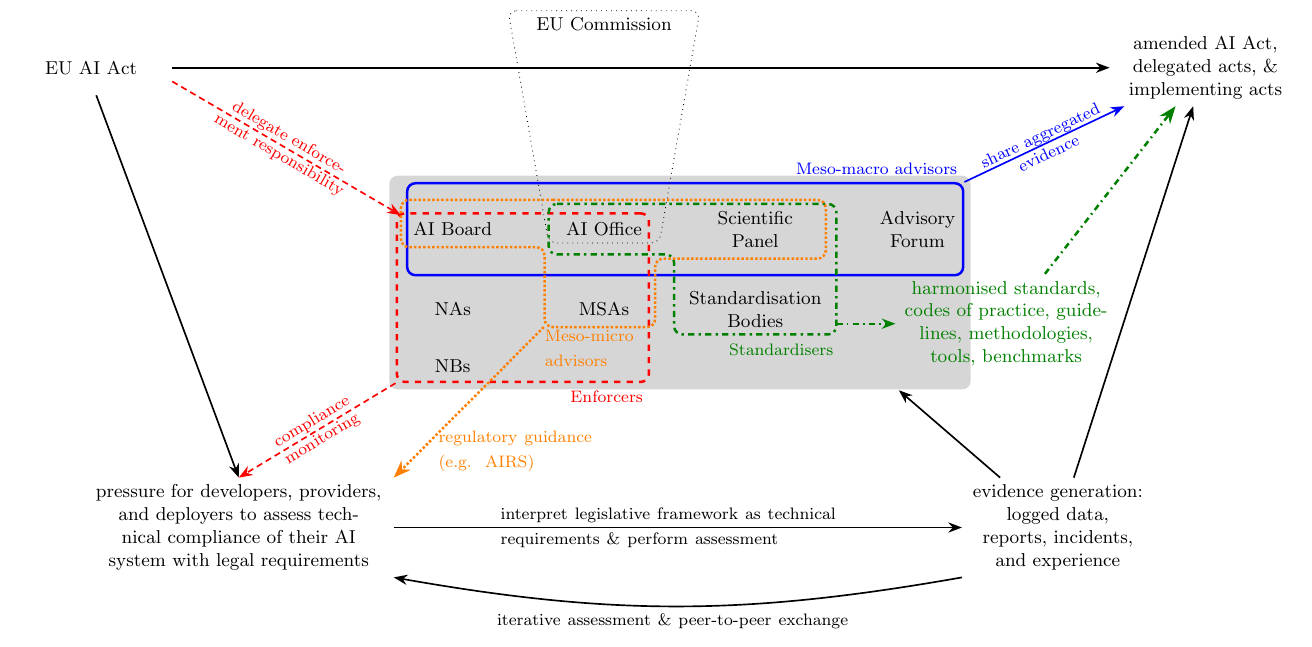}
    \caption{The EU AI Act's bathtub.}
    \label{fig:bathtub}
\end{figure*}

\subsection{The AI Act's Bathtub}

Applying the previously established functional lens to the actors in the AI Act's learning landscape allows mapping them, based on their tasks in both enforcement and learning, to an abstract three-level framework constituted of the micro, meso, and macro levels.
This idealised mapping is shown in Figure~\ref{fig:bathtub}.

At the source of the regulatory learning pipeline is the EU AI Act itself, providing the legislative framework that exerts pressure on the lowest level by demanding technical compliance with the legislation.
This responsibility of demonstrating technical compliance with legal requirements belongs to providers, deployers, and developers of AI systems, including large firms, \acp{SME}, startups, and even public institutions.
However, these organisations may be supported by initiatives such as \acp{EDIH} or \acp{TEF} and even members of civil society, where the system is open to third-party scrutiny.
Thus, they form the micro level of the framework.
The pressure exerted on them by the macro level may cause learning at the micro level itself, as the actors affected at this level need to navigate the legal uncertainty and adjust their design, development, and assessment practices accordingly.
These changes then constitute the valuable learning signals and serve as the micro-level evidence that can be aggregated upward and, in turn, enable data-driven adaptation of the macro level, i.e.\, through direct amendments of the AI Act, delegated acts, or implementing acts.
As the European Commission is the sole actor with the legal power to exert these adaptations, the Commission represents the macro level.

In contrast to the micro and macro levels, the meso level consists of those actors who are tasked with the translation of the macro-micro transition on the one hand and evidence aggregation of the micro-macro transition on the other, effectively constituting the mediation filter between the macro and micro actors.
Considering the meso actors in terms of translation, in the form of enforcement or guidance, and aggregation, creates four different functional categories that the actors can be grouped into.
As the example of the AI Office shows, the AI Act assigns multiple tasks to each actor. Therefore, we consider an exclusive categorisation of an actor into a single category to be not possible.
Nevertheless, the categorisation aims at providing clarity in discussing the role of each actor along the enforcement and learning axes.

The first of these categories consists of the \textit{Enforcers}, actors delegated with enforcement of the legal requirements.
This category includes, at a national level, the \acp{CA}, i.e., the \acp{MSA} and \acp{NA}, as well as \acp{NB}.
Together, they are tasked with monitoring and assessing conformity of the micro level actors with the legal requirements laid out by the AI Act.
It should be noted that \acp{NA} can be characterised as indirect enforcers, in the sense that their enforcement role is exercised primarily through the designation, appointment, and oversight of \acp{NB}, which in turn carry out third-party conformity assessments of AI systems.
The group is completed by supranational bodies in the form of the AI Board and the AI Office.
The former, composed of Member State representatives, assumes an enforcing role through the coordination of the national \acp{CA}, whereas the latter is tasked with direct monitoring of \ac{GPAI} models and supporting the AI Board administratively as secretariat.
Collectively, these actors exert the necessary top-down pressure on the micro level, ensuring that the theoretical obligations of the Act are met with actual technical compliance.

The macro-micro transition does not, however, solely consist of enforcement activities.
Instead, some actors defined by the AI Act are tasked with supporting activities through guidance and help in translating the legislation into technical requirements.
These actors, notably the AI Board, the AI Office, and \acp{MSA}, constitute the \textit{Meso-micro advisor} category.
In addition, the Scientific Panel is mandated to provide external expertise to \acp{MSA} at their request, thus playing a crucial role in e.g., \acp{AIRS}, but also in supporting the AI Office in its tasks.
As such, they primarily act in a meso-meso advisory role.
Their meso-micro advisory function is to provide direct operational support to the micro level, specifically through the coordination of administrative practices and the facilitation of \acp{AIRS}.
By managing these \acp{AIRS} and ensuring consistent application of rules across Member States, they create a controlled environment where developers can experiment and receive immediate regulatory feedback.

In the micro-macro transition, the majority of the meso-level actors serve as the crucial aggregators of the micro-level evidence, acting as a filter for the signals enabling policy adaptation.
The first category consisting of these aggregators is the \textit{Meso-macro advisors}.
These actors receive the evidence produced by the micro level and the enforcers to then pass it on to the macro level, serving as the primary feedback loop for the \ac{EU} Commission.
This category includes the Advisory Forum, which aggregates stakeholder perspectives, as well as the AI Board and AI Office in their reporting capacity.
By synthesising raw data, such as incident reports, market trends, or \ac{AIRS} outcomes, into structured opinions and recommendations, they provide the Commission with the necessary signal to trigger regulatory learning and adaptation.

Lastly, the \textit{Standardisers} category groups those actors who translate the legislative provisions into concrete technical and regulatory specifications.
The major actors in this category are the European Standardisation Bodies, such as \ac{CEN}, \ac{CENELEC}, and \ac{ETSI}.
They develop standards, including those at the request of the \ac{EU} Commission, that then become regulatory instruments in the form of harmonised standards and ultimately serve as the key instruments for sector-specific, vertical governance by presuming conformity with the AI Act.
However, this functional category also extends to actors such as the AI Office when they develop codes of practice and implementation guidelines.
Like technical standards, these instruments serve to bridge the gap between abstract legal commands and practical application, providing the detailed specifications necessary for the micro level to demonstrate conformity.
While serving as intermediary between legislation and technical requirements, these instruments become tools in the bottom-up feedback to assess the feasibility and potentially adapt the legislation based on the micro-level experience.

Nevertheless, the bathtub model in Figure~\ref{fig:bathtub} does not aim to represent all interactions between the mapped actors, in the interest of visual clarity. Even with this abstraction, the categorisation of meso-level actors into the four outlined categories illustrates the structural complexity of the AI Act’s collaborative governance approach.
Considering the interplay of actors further reinforces this complexity, as individual actors may assume multiple roles, while specific roles may be shared across multiple actors. 
In particular, while the AI Act explicitly assigns certain tasks that prioritise the input of the AI Board and the AI Office at the macro level, for example relative to standardisation bodies, these assignments do not amount to a uniformly hierarchical ordering across all actors.
Notably, national \acp{NB} constitute the clearest instance of formal subordination, as their operation is contingent on designation and oversight by the \acp{NA}.
Taken together, this observation shows the importance of well-defined and coherent evidence generation at the micro level.
Such coherence is a prerequisite for interoperable analysis across meso-level actors and, ultimately, for the effective aggregation of evidence necessary to support regulatory learning.

\section{Boundary Objects and Negotiating Artifacts}
\label{sec:boundaries}
The bathtub model established in the previous section provides an idealised account of the regulatory pressure, the origin of learning, and the mediation between the macro and the micro levels.
It identifies the micro level as the site at which the burden of generating learning signals is concentrated, and where the legal uncertainty caused by abstract requirements is most directly encountered.
The model further visualises the diverse interactions between micro and meso actors, whose differing governance roles entail diverging perspectives on the same task of evaluating compliance of a technical system with codified ethical values. 
As such, the bathtub provides an architectural account of the \textit{where} and the \textit{who} of the regulatory learning cycle, abstracting away the messy real-world interactions between actors.
Hence, it presents analytical limitations in explaining the mechanism through which these actors mediate legal uncertainty into meaningful upward signals, and therefore leaves open the \textit{how}.
This gap becomes especially visible upon detailed consideration of the interactions between multidisciplinary stakeholder groups involved in translating abstract legal requirements into operational practice, which is where the notions of \textit{boundary objects} and \textit{\acp{BNA}} may be analytically useful.

Introduced by \citeauthor{starInstitutionalEcology1989}, the concept of boundary objects captures artifacts ---abstract or concrete--- that occupy the space between distinct communities of practice~\cite{starInstitutionalEcology1989, leighstarThisNotBoundary2010}.
For an artifact to qualify as a boundary object, it requires a certain degree of robustness to preserve a shared core understanding and an adequate degree of plasticity to accommodate the information needs of different actors~\cite{starInstitutionalEcology1989, leeBoundaryNegotiatingArtifacts2007}.
In socio-technical settings, this is particularly important as coordination between distinct stakeholder groups requires a shared reference point, which is the boundary object in question~\cite{beddoesUsingBoundaryNegotiating2011a}.
In AI governance, the same AI system may be simultaneously a technical artifact to be engineered, a legal object to be assessed, and an organisational object to be managed, while the applicable requirements may be treated as a specification, an obligation, or a benchmark.
The former thus acts as the technical boundary object, with the latter representing the regulatory boundary object.

\acp{BNA} expand on the notion of boundary objects by specifying the concrete artifacts and practices through which such shared reference points are rendered useful in coordinated collaboration~\cite{leeBoundaryNegotiatingArtifacts2007,leeEmbracingChaosAgain2026}.
Where boundary objects identify a common reference point across communities, \acp{BNA} help complete the framework of a coordinative space by emphasising the work required to produce, maintain, and iteratively stabilise meanings across boundaries.
This includes the documents, models, and processes used by any given community of practice to communicate their perspective to another community~\cite{leeBoundaryNegotiatingArtifacts2007}.

As \citeauthor{leeBoundaryNegotiatingArtifacts2007} points out, the original account of boundary objects given by \citeauthor{starInstitutionalEcology1989} relies on standardisation either through the established or naturally pre-existing structure of an object, such as the distinct characteristics of a bird~\cite{leeBoundaryNegotiatingArtifacts2007}.
By contrast, \acp{BNA} are especially useful when such an inherent structure is absent or insufficient.
Then, the coordination itself has to be actively negotiated through collaborative practice.
This has been explored in inter- and multidisciplinary team settings, where \acp{BNA} help explain how heterogeneous actors align around shared tasks without requiring full agreement on interpretation~\cite{beddoesUsingBoundaryNegotiating2011a}.
In this study, the authors show that artifacts such as literature reviews, wikis, tables, charts, and summary reports served as practical means of proposing ideas and sustaining shared understanding of complex work.
Similarly, in engineering design contexts, prototypes and related artifacts can function as interfaces through which actors negotiate requirements, evaluate alternatives, and stabilise provisional understandings across phases of development~\cite{subrahmanianBoundaryObjectsPrototypes2003}.

Such multi-stakeholder negotiation, while not explicitly labelled in boundary negotiation terms, has been the subject of the extant literature on policy prototyping.
Notably, specially designed workshops organising stakeholder collaboration have experimented with abstract legal policies and attempted to turn them into concrete operational requirements~\cite{kimbellPrototypingNewSpirit2017,mcnealyPrototypingPolicyVisualizing2022}.
This approach resembles experimental policy-making insofar as it creates prototypes of the high-level policies and then tests them in controlled environments~\cite{nunoOpenLoopPolicyPrototyping2022}.
In their report, \citeauthor{nunoOpenLoopPolicyPrototyping2022} analyse the results from the policy prototyping activities they engaged in with over fifty companies that collaborated via the online platform \textit{Open Loop Forum}~\cite{nunoOpenLoopPolicyPrototyping2022}.
Identifying issues with the clarity and feasibility of requirements, including risk management, transparency, and human oversight, the workshop produced actionable recommendations for improving the regulation, such as an increased focus on collaboration for \acp{AIRS}.
The requirement for human oversight was the subject of another policy prototyping workshop led by \citeauthor{oomsPolicyPracticePrototyping2025}~\cite{oomsPolicyPracticePrototyping2025}.
In their study, a diverse set of stakeholders, from companies to civil society to technical experts, engaged in four phases that ranged from mapping the use cases and technical properties to the drafting of compliance documents as concrete prototypes to verify the operationalisation of the legal framework.
Legal uncertainty, caused by vague legal requirements, was identified as problematic, but participants noted the advantage of flexibility allowing multiple stakeholder perspectives to shape the prototypes, ultimately benefitting the ecosystem.

However, the organisation of such dedicated workshops raises questions as to the scalability of regulatory learning.
In contrast, the compliance obligations of the \ac{AI Act} give rise to three distinct, but overlapping and naturally occurring settings in which boundary negotiation happens and may produce valuable signals for regulatory learning.
First, the \textit{a priori} simplest scenario in terms of involved actors and their interactions is the conformity assessment of a high-risk AI system based on internal controls (Article 43(1)(a) and 43(2), Annex VI).
This could, for instance, be an \ac{SME} developing an AI-based software to evaluate the suitability of candidates based on their profile, qualifying as high-risk by Article 6(2) and Annex III of the AI Act, therefore requiring compliance with Articles 8 to 27.
In this instance, the regulatory learning would mostly remain self-contained at the micro level as the \ac{SME}, through internal stakeholder deliberation, translates the legal requirements into concrete design and assessment methodologies and metrics for the system they develop.
As the \ac{SME} progresses through the life cycle of its AI system, iterations of the self-assessment may yield continuous improvement of their legal understanding, as every interpretative choice made may reduce the legal uncertainty.

Second, the \ac{AI Act} provides for a supervisory, and therefore collaborative, framework through Article 57 in the form of \acp{AIRS}.
Here, a designated \ac{MSA} offers legal guidance, technical expertise, and technical resources to an applying provider or deployer.
This scenario, codified in the legislation, thus represents a formal channel between the micro and the meso level.
The third scenario, namely the conformity assessment by a third party such as a notified body, may be seen as a continuation of either of the two previous settings.
In fact, the third-party conformity assessment consists in an audit of the technical documentation produced by the provider or deployer, as per Articles 43(1)(b), 43(3), and Annex VII.
With respect to the regulatory learning cycle, the difference from the \ac{AIRS} setting is the intermediary aggregation of information by the third-party assessor before further communication with the competent \ac{MSA}.
While all three settings are relevant, we will focus in the following on the \ac{AIRS} due to its privileged formal status as an explicitly structured setting for regulatory learning.

\section{AI Regulatory Sandboxes are Environments of Boundary Negotiation}
\label{sec:AIRS}
Among the multitude of provisions implementing adaptive and experimental legislation in the \ac{AI Act}~\cite{schrepelAdaptiveRegulation2025a}, \acp{AIRS} have been of particular focus to scholars, with \citeauthor{ahernNewAnticipatoryGovernance2025} highlighting their importance, but noting the friction that the fresh mindset of anticipatory governance creates~\cite{ahernNewAnticipatoryGovernance2025}. 
Articles 57 to 59 of the \ac{AI Act}~\cite{EUAIAct2024}, in addition to the draft implementing act for \acp{AIRS}~\cite{EUComDraftAIRS2025}, outline several objectives, notably innovation support, support for conformity, and legal experimentation.
The former two have been the subject of a deeper analysis by \citeauthor{ahern2025Operationalising}, who notes risks of fragmentation in Europe and lack of incentives for prospective participants~\cite{ahern2025Operationalising}.
Moreover, the exact provisions align most closely with the goal of legal guidance, as difficulties in testing in real-world conditions and interactions with other digital regulation are uncertain and underspecified~\cite{genicotExploringBoundariesAI2025}.

The challenges associated with the testing phase of \acp{AIRS} are recognised more widely with respect to the legal uncertainty, with the potential for its reduction discussed as an important criterion for admitting a participant to an \ac{AIRS}~\cite{novelliGettingRegulatorySandboxes2026}.
This represents a shift in rationale from the more established narrative of supporting innovation, aligning with design considerations put forward by \citeauthor{chenDesigningRegulatorySandboxes2026} who differentiate between exploratory, confirmatory, and on-demand regulatory sandboxes~\cite{chenDesigningRegulatorySandboxes2026}.
\acp{AIRS}, during times of significant legal uncertainty for both practitioners and regulators, should therefore be primarily of the exploratory type, before gradually becoming confirmatory as the regulation stabilises.
Nevertheless, \citeauthor{chenDesigningRegulatorySandboxes2026} primarily consider the regulator's role as monitoring, even in exploratory settings, with the onus of experimentation falling onto the providers.
Similarly, \citeauthor{guioespanolRegulatorySandboxesAI2025} consider \acp{AIRS}' potential for accelerated learning towards reducing legal uncertainty~\cite{guioespanolRegulatorySandboxesAI2025}.
In contrast to the other works, the authors specifically stress the need for mutual learning and robust collaboration between stakeholders in order to achieve the stated goals.

\begin{figure*}
    \centering
    \includegraphics[width=0.85\linewidth]{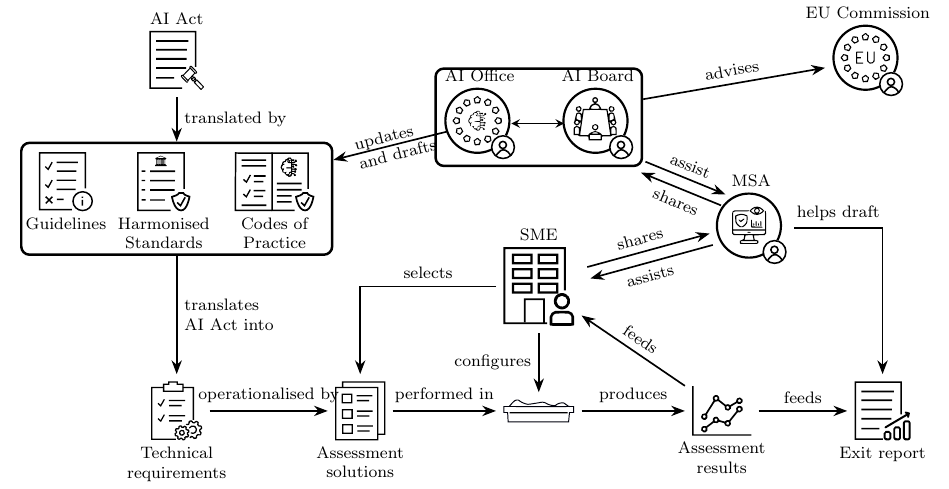}
    \caption{Idealised flow of information in an \ac{AIRS} and its feedback loop in regulatory learning.}
    \label{fig:AIRS}
\end{figure*}

From the above considerations, \acp{AIRS} are therefore a site in which the meaning of legal requirements is actively established through interaction among \ac{AIRS} participants.
A detailed description of the low-level workflow of an \ac{AIRS}, from the application of a company to the joint drafting of an exit report can be found in~\cite{buscemi2025sandbox}.
An idealised, higher-level flow situating an \ac{AIRS} engagement in the regulatory learning cycle, from top-down pressure to bottom-up learning, is shown in Figure~\ref{fig:AIRS}.
The driving stimulus remains, as has been the assumption throughout, the legal pressure exerted by the \ac{AI Act}.
In the absence of planned standards and other related regulatory instruments, the translation of the legal into technical requirements and their subsequent application by appropriate assessment solutions remains at the discretion of the \ac{SME}.
Due to their participation in the \ac{AIRS}, however, they receive assistance from the competent \ac{MSA}.
For instance, the \ac{SME} may be developing an AI system for recruitment purposes, which would qualify as high-risk per Annex III of the \ac{AI Act}.

What the \ac{AIRS} adds in such a setting is a structured arena for boundary negotiation.
The AI system under development and the applicable legal requirements become, as mentioned, the two primary boundary objects around which the involved actors coordinate.
On the one hand, the \ac{SME} must translate abstract legal requirements into feasible design choices, assessment procedures, and evidence forms.
On the other hand, the \ac{MSA} contributes legal interpretation, administrative experience, and potentially technical expertise.
Beyond legal guidance and supervisory support, the deliberation over boundary objects especially generates learning signals that are shaped by the system's socio-technical dimensions, which ultimately depend on multi-stakeholder perspectives.

Even where standards and regulatory instruments are available, technological development is likely to outpace their specification, leaving grey areas that cannot be fully resolved ex ante.
\acp{AIRS} are therefore particularly valuable as the first structured environment in which such disruptive systems can be examined under regulatory supervision and multi-stakeholder collaboration.
This becomes especially relevant when novel technologies fall partly outside the scope of existing frameworks, as recent work on AI agents has suggested~\cite{nanniniAIAgentsEU2026,yousefiAgenticAIEU2025}.

Under such conditions, the \ac{AIRS} does not simply support the application of already stabilised requirements, but becomes a setting in which such an interpretation has to be worked out through \emph{iterative} boundary negotiation.
As the \ac{SME} refines its system and the corresponding assessment methodology, and the \ac{MSA} responds to these choices, a provisional stabilisation of meaning may emerge around what counts as, e.g., sufficient robustness, fairness, documentation, or human oversight.
The \ac{AIRS} therefore functions as more than a channel of top-down assistance.
It creates an institutional environment in which bottom-up learning signals may be generated and significantly shaped by the participants through iterated negotiation. 

As the EU Commission draft for implementing \ac{AIRS} points out in paragraph 21, \enquote{[...] dissemination of the results of a specific sandbox project for the benefit of everyone should encourage dialogue among and between competent authorities, participants and non-participants, such as relevant stakeholders}~\cite{EUComDraftAIRS2025}.
The exit report jointly drafted by the \ac{AIRS} participants thus is not only an artifact that serves as an audit trail of the \ac{AIRS} activities and micro-level learning.
As it is shared with the relevant stakeholders, the exit report constitutes a further \ac{BNA} that may be used at the meso level for an analysis of trends in operationalising the legal requirements.
In this sense, the \ac{AIRS} extends beyond the assessment of an individual system.
It does so through the production of such artifacts, creating the possibility that case-specific interpretation may become reusable beyond the immediate interaction between regulatory sandbox participant and competent authority.

\section{Technical Frameworks for Boundary Negotiation}
\label{sec:BNA-sandbox}
As other scholars have pointed out, the legal framework for \acp{AIRS} is insufficient to support meaningful regulatory learning on its own.
To fulfil their purpose of generating significant learning signals, they also require the appropriate technical infrastructure, an essential component that remains underdeveloped~\cite{genicotExploringBoundariesAI2025, ahern2025Operationalising}.
In \cite{lewis2025regulatorylearning}, \citeauthor{lewis2025regulatorylearning} build an extensive inventory of the actors, their responsibilities, and activity-level information flow of the \ac{AI Act}'s regulatory learning space.
To combat the legal uncertainty caused by codified fundamental rights and associated standards, they propose the use of open linked data-based semantic frameworks to provide a technical backbone capable of structuring the information exchange.
However, such an approach concerns solely the data exchange and does not by itself serve as an artifact structuring boundary negotiation between diverse stakeholders.
In this section, we therefore analyse two technical frameworks, proposed in the literature, as illustrations of how boundary negotiation may be operationalised in practice and serve as technical infrastructure in support of \acp{AIRS}.

Building on their analysis of the actors and activities of \acp{AIRS}, \citeauthor{buscemi2025sandbox} extract requirements for potential technical infrastructure, i.e., \acp{AITS} \cite{buscemi2025sandbox}.
Given these requirements, they further propose the \textit{Sandbox Configurator}, a technical framework designed to support the flexible instantiation of \acp{AITS} for the technical assessment of AI systems in an \ac{AIRS}.
The core of the framework consists in its proposed \ac{DSL}, a declarative machine-readable specification language for the configuration of an \ac{AITS} as a structured technical assessment environment.
As the authors note, \enquote{[t]he \ac{DSL} serves as a contract between stakeholders}, with encoding of compliance requirements, testing objectives, and domain-specific modules, all provided by different stakeholder groups.
The encoding of these specifications in a machine-readable, but declarative format makes it interesting for the \textit{Sandbox Configurator} to function as a \ac{BNA}.
In fact, it is particularly useful given that the design, development, and assessment of the AI system constitute an iterative process, such that the traceability afforded by explicit encoding of specifications can reveal the stabilisation of the \ac{BNA} once changes over iterations become minimal. 
Moreover, it allows the identification of divergences and agreements between stakeholder groups, providing a multi-dimensional view of the assessment activities to all actors involved.

In addition to encoding stakeholder perspectives and decisions in the \ac{DSL} of the \textit{Sandbox Configurator}, the framework provides role-specific access to the assessment environment via visual dashboards, granting it communicative capabilities that are essential for \acp{BNA}.
This perspective adaptation can support the flexibility needed from a \ac{BNA}, as it may present relevant information tailored to the needs of certain stakeholder groups.
\citeauthor{ramliReferenceArchitecturesBoundary2025} found that boundary objects, and by extension therefore also \acp{BNA}, in the form of reference architectures, can create synergy between collaborating stakeholders, bring about change, and entail learning~\cite{ramliReferenceArchitecturesBoundary2025}.
Last but not least, the framework's container-based deployment allows for flexible instantiation of these \acp{AITS} on various infrastructures, ultimately providing easier setups for collaboration as stakeholders could potentially engage remotely with the assessment environment in a federated setting.

A second relevant proposal studying multi-stakeholder engagement in \acp{AIRS}, in the public sector specifically, was made by \citeauthor{gonzaleztorresRoleRegulatorySandboxes2023}. 
In their paper, the authors propose the adoption of \ac{MLOps} paradigms, methodologies, and infrastructure to serve as the operational backbone of \acp{AIRS}, making regulatory experimentation administratively and technically workable~\cite{gonzaleztorresRoleRegulatorySandboxes2023}.
By referring to \textit{virtual} sandboxes, their contribution explicitly frames \ac{AIRS} as an operational environment in which different actors collaborate through shared technical artifacts and procedures.
The \ac{MLOps} approach is particularly relevant because it structures the design, development, and assessment process into stages across the AI system life cycle, while also supporting role-based access to the same underlying technical infrastructure.
This allows different stakeholder groups to engage with the same artifact according to their information needs, while preserving traceability and auditability through the outputs produced at each stage.
In particular, the distinction between core stakeholders and additional participants such as standardisation bodies or civil society reflects the way \ac{MLOps} can support boundary negotiation.
The workflow remains sufficiently stable to coordinate action, yet flexible enough to accommodate different perspectives on the boundary objects.
Through repeated interaction with the pipeline and its outputs, stakeholders can progressively refine the system, its requirements, the interpretation of regulatory requirements, and assessment practices, thereby contributing to the stabilisation of the \ac{BNA} and the boundary objects.

A brief synthesis of the two technical frameworks discussed in this section suggests several recurring properties of technical \acp{BNA} for AI assessment.
They require a combination of a machine-readable core, role-specific access, and iterative outputs that can be reused across assessment cycles.
In the two frameworks discussed here, that means a shared \ac{DSL} or pipeline that can encode requirements and testing objectives, dashboards or staged access that adapt the same artifact to different stakeholder groups, and traceable results such as logs, assessment results, and exit reports that make negotiation visible over time.
Such frameworks support the gradual stabilisation of assessment practices across iterations and stakeholder groups.

\section{Discussion}
\label{sec:discussion}
As discussed in Section~\ref{sec:ai-act-bathtub}, the collaborative governance structure defined by the \ac{AI Act} reveals a complex network of actors with multi-dimensional roles.
Applying Coleman's bathtub as an analytical model carries several normative implications that are important for operationalising the learning cycle in this space.
In fact, every actor has defined responsibilities, with annual reviews, workshops, and \acp{AIRS} constituting formal mechanisms for regulatory learning.
The bathtub model that arises from these definitions allows the clear identification of the micro level and its primary task of evidence generation ---beyond its need to demonstrate compliance--- and the meso level for mediation and aggregation.

To better understand the production of learning signals, we have further interpreted an AI system in its socio-technical context and corresponding legislation as the main boundary objects, and supporting technical frameworks as the corresponding \acp{BNA}.
This framing is particularly useful in moving away from a technocratic perspective in which the only evidence considered as useful is the quantitative metrics and threshold-based criteria.
Instead, it focuses on the sociological processes involved and not only their dependence on shared reference points, but on the negotiation work required to stabilise these reference points and keep them flexible for continued interpretation.
Only then do technical frameworks that enable multi-stakeholder and interdisciplinary collaboration, like the previously considered \textit{Sandbox Configurator}, truly enable boundary negotiation through their consolidation of evidence and malleable presentation layer.

However, the understanding of processes and artifacts through boundary objects and \acp{BNA} may be useful beyond the application of a sociological lens to an environment of multi-stakeholder collaboration and boundary negotiation shaped by legal uncertainty.
It highlights a key risk in rushing towards premature drafting of regulatory instruments such as standards in attempts at providing legal certainty as fast as possible.
As \citeauthor{schiffStrategiesHarmonizingFragmented2025} points out, the landscape of standards is fragmented~\cite{schiffStrategiesHarmonizingFragmented2025}.
Premature stabilisation or closure of the boundary objects, through over-specification, would preclude adaptability of the drafted regulatory instruments to advancing developments in AI systems, thus requiring the need for further specifications, causing increased fragmentation.

These risks of regulatory capture are highlighted in the institutional analysis by \citeauthor{cancela-outedaEUsAIAct2024}, who acknowledges that the governance structure does foster stakeholder collaboration and input at the regulatory level, but warns against abuse of power by certain stakeholder groups~\cite{cancela-outedaEUsAIAct2024}.
As regulatory learning is inherently political, it is subject to significant influence from the interests of the involved stakeholders and decision makers.
Such risks are confirmed by power dynamics complicating boundary negotiation~\cite{leeEmbracingChaosAgain2026,hawkinsBoundaryObjectsPower2017}.
At the legislative level, this manifests as a question of political willingness.
This issue is further compounded at the meso level where the data are aggregated and filtered before being passed to the macro level, especially in the standardisation efforts intended to provide vertical governance.
As noted by Metcalf et al.~\cite{metcalfAISafetyRegulatory2025}, standardisation processes often involve stakeholders with asymmetric resources and capacities, which creates a non-negligible risk of regulatory capture.
The researchers identify that, in such contexts, the interpretation of legal requirements into technical operationalisations may tend to prioritise feasibility, scalability, or market compatibility, potentially at the expense of more robust protections of fundamental rights.
The interests of the micro-level actors in particular, such as \acp{SME}, should be given more weight~\cite{kilianEuropeanAIStandards2025}, a step that the AI Act has already taken through their inclusion in the regulatory learning process via the Advisory Forum and their priority in \ac{AIRS} admissions.
As \citeauthor{guioespanolRegulatorySandboxesAI2025} point out, the asymmetry in technical expertise between technical and regulatory stakeholders can lead to undermining of regulation~\cite{guioespanolRegulatorySandboxesAI2025}.
In this instance, previous analyses of \acp{BNA} and boundary negotiation, such as those by \citeauthor{leeEmbracingChaosAgain2026}, provide a means of analysing these asymmetries and designing \acp{BNA} that minimise the risk of asymmetry exploitation~\cite{leeEmbracingChaosAgain2026} at the micro level.

The dangers of regulatory capture in boundary object standardisation thus become dual, in the sense that power dynamics may be used by certain stakeholders, causing a short-circuiting of the feedback loop, premature closure of the boundary objects, and checkbox compliance behaviour instead of fruitful interpretative deliberation.
As \citeauthor{schiffEmergenceArtificialIntelligence2024} point out in their analysis of auditing practices in AI compliance, the active stakeholder collaboration amid legal uncertainty was a major driver in this community of practice's sizeable contributions to the ecosystem in the form of methodologies, processes, and technical assets~\cite{schiffEmergenceArtificialIntelligence2024}.

It has to be acknowledged that legal uncertainty cannot be fully removed ex ante. At the same time, framing it as a temporally bounded, productive condition for regulatory learning is not an endorsement of uncertainty for its own sake and should not be used as justification for intentional under-specification of requirements.
The implication is that a limited degree of residual uncertainty creates the space in which boundary negotiation, iterative assessment, and regulatory learning can take place before more stable interpretations emerge.
The aim of adaptive regulation, whether at the macro level through the legislation itself or through appropriate soft regulatory instruments like standards, needs to remain aimed at guaranteeing legal certainty for all stakeholders, from developers to policymakers and auditors.
Understanding regulations as boundary objects, however, suggests that this certainty is best achieved through flexible specification rather than premature closure.
Scalable and reproducible evidence generation through technical frameworks, such as consistent methodologies in technical sandboxes, becomes increasingly important for this purpose, as these data would allow the identification of common metrics, practices, and methods across operationalising settings.
Having a common core, without over-specification of detailed metrics and associated thresholds, may therefore increase long-term legal certainty while preserving flexibility for meaningful boundary negotiation of the socio-technical interpretation of the legal requirements and adaptation to specific use cases.
This flexibility on top of a robust, malleable core corresponds to \citeauthor{schiffStrategiesHarmonizingFragmented2025}'s strategy for harmonising standards with flexible, yet horizontal specifications that allow for sectoral and use case-specific adaptations~\cite{schiffStrategiesHarmonizingFragmented2025}.

Last but not least, the boundary negotiation-based analysis carries implications for the role and design of \acp{AIRS}.
Intended as a measure in support of innovation, the role of the \ac{AIRS} for participants should not just be the provision of \enquote{increased legal certainty} by the competent authority.
Instead, in its intention to also be an environment of experimentation and for enabling regulatory learning, it must actively encourage the boundary negotiation from all participants.
This renders the appropriate technical infrastructure and its requirements increasingly important as technical and legal stakeholders contend the meaning.
As the EU Commission's EUSAiR project shows in its pilots, continued efforts are already being put into the design of \textit{digital} sandboxes to provide a technical backbone to the legal framework~\cite{eusair_roadmap_2025}.

In addition to offering analytical usefulness through its socio-technical framing, the present paper is also subject to a handful of limitations.
First, the study focused on the EU \ac{AI Act} as a case study, potentially limiting its potential for generalisation to other jurisdictions.
However, the Brussels effect, where regulatory spaces outside of the \ac{EU} adopt \ac{EU} legislation, at times through voluntary application by the industry, indicates that analogous considerations may still apply~\cite{bradfordBrusselsEffect2020, ebersTrulyRiskbasedRegulation2025, gstreinGeneralpurposeAIRegulation2024}.
There are early indications of other jurisdictions already adopting a similar, even if less formal approach to the establishment of adaptive regulation, such as the Artificial Intelligence Basic Plan in Japan~\cite{ArtificalIntelligenceBasicJapan2025} or the soft-governance approach in South East Asia~\cite{putraGoverningAISoutheast2024a}.
Hence, the methodology of mapping actors according to the bathtub can be adapted to the precise provisions of those frameworks.
The voluntary, standards-based compliance regulation of AI systems in the United States makes the underlying discussions on the stabilisation of boundary objects and related standard drafting relevant for the American market.
Moreover, the underlying considerations of boundary objects and negotiations remain relevant as AI systems continue to pose socio-technical challenges, even if other frameworks are less explicitly fundamental-rights focused as the \ac{AI Act}.

Second, the notions of boundary objects and \acp{BNA} have been developed in specific collaborative settings, with their applicability in regulatory contexts still to be determined.
To our knowledge, we are the first to apply this lens in this context and believe that the socio-technical nature of AI regulation warrants future investigations into such an interdisciplinary analysis.

Third, we limited our discussion of technical frameworks as \acp{BNA}, specifically in support of \acp{AIRS}, to only two illustrations.
However, other notable solutions and frameworks exist that deserve consideration in how they can help with boundary negotiation and subsequent regulatory learning in the policy development for AI governance.
The PARMA architecture proposed in~\cite{pintzPARMAPlatformArchitecture2024} for instance, provides flexible and modular workflow definition for the development and assessment of AI systems with multi-stakeholder collaboration in mind.
Similar solutions have also been introduced in industrial settings with the explicit goal of ecosystem development and standard-drafting support, such as the sandbox initiative led by the AI Verify Foundation in Singapore~\cite{GlobalAIAssurance}.

Lastly, while the aforementioned solutions directly support the micro-level learning and stabilisation of boundary objects, they are on their own limited in enabling meaningful aggregation and analysis of these learning signals at the higher levels.
To enable evidence sharing, a combination of these frameworks with other solutions, such as the semantic frameworks and their machine-readable implementations mentioned above and in~\cite{lewis2025regulatorylearning, golpayeganiAICardsApplied2024a,golpayeganiAIROOntologyRepresenting2022} present an open design challenge for future work and ultimately a way towards scalable, operational regulatory learning.

\section{Conclusion}
\label{sec:conclusion}
This paper has moved the focus on the \ac{AI Act}’s regulatory learning space from a purely legal structure to that of a socio-technical process in which abstract requirements are translated into operational practice through interaction across micro, meso, and macro levels.
Using Coleman’s bathtub, it mapped how regulatory pressure, evidence generation, and aggregation connect these levels, and argued that this process depends, amid persisting legal uncertainty, on iterative multi-stakeholder boundary negotiation.
The analysis further suggests that legal uncertainty cannot be fully reduced ex ante and that, within reasonable bounds, it is also necessary for regulatory learning, because it creates the interpretative space in which stakeholders negotiate and progressively stabilise the meaning of abstract legal requirements. 
Regulatory sandboxes and related technical frameworks can support this negotiation by providing shared artifacts, traceable outputs, and role-specific access to the same assessment environment.
In addition, the analysis of these frameworks identifies concrete properties that technical \acp{BNA} require in order to support AI assessment effectively.
The challenge, therefore, is not to eliminate uncertainty immediately, but to structure it in ways that enable reusable evidence, iterative learning, and eventual legal certainty without premature closure.

\appendix

\section*{Acknowledgments}
We thank the Luxembourg AI Factory (L-AIF, grant agreement ID 101234366) for supporting this work.

% \bibliography{refs}
\bibliography{refs_bathtub}


\acrodef{AI}{Artificial Intelligence}
\acrodef{FCA}{Financial Conduct Authority} 
\acrodef{AI Act}{Artificial Intelligence Act}
\acrodef{GDPR}{General Data Protection Regulation}
\acrodef{CRA}{Cyber Resilience Act} 
\acrodef{SME}{Small and medium-sized enterprise}
\acrodef{GPAI}{General Purpose AI}
\acrodef{DGA}{Data Governance Act}
\acrodef{DSA}{Digital Services Act}
\acrodef{DMA}{Digital Markets Act} 
\acrodef{AIRS}{AI Regulatory Sandbox}
\acrodefplural{AIRS}[AIRSes]{AI Regulatory Sandboxes}
\acrodef{AITS}{AI Technical Sandbox}
\acrodefplural{AITS}[AITSes]{AI Technical Sandboxes}
\acrodef{MSA}{Market Surveillance Authority}
\acrodefplural{MSA}{Market Surveillance Authorities}
\acrodef{EUSAiR}{EU Regulatory Sandboxes for AI}
\acrodef{LIST}{Luxembourg Institute of Science and Technology} 
\acrodef{CNPD}{National Data Protection Commission}
\acrodef{EDIH}{European Digital Innovation Hub} 
\acrodef{TEF}{Testing and Experimentation Facility}
\acrodefplural{TEF}{Testing and Experimentation Facilities}
\acrodef{EuroHPC}{European High-Performance Computing}
\acrodef{EDIC}{European Digital Infrastructure Consortia}
\acrodef{AIoD}{AI-on-Demand Platform}
\acrodef{AI RMF}{AI Risk Management Framework}
\acrodef{CA}{Competent Authority}
\acrodefplural{CA}{Competent Authorities}
\acrodef{NA}{Notifying Authority}
\acrodefplural{NA}{Notifying Authorities}
\acrodef{NB}{Notified Body}
\acrodefplural{NB}{Notified Bodies}
\acrodef{NIST}{National Institute of Standards and Technology}
\acrodef{RMF}{Risk Management Framework}
\acrodef{CEN}{European Committee for Standardisation}
\acrodef{CENELEC}{European Committee for Electrotechnical Standardisation}
\acrodef{JTC}{Joint Technical Committee}
\acrodef{RAG}{Retrieval Augmented Generation}
\acrodef{NLP}{Natural Language Processing}
\acrodef{HPC}{High Performance Computing}
\acrodef{RISE}{Research Institutes of Sweden AB}
\acrodef{HLEG}{High-level Expert Group on Artificial Intelligence}
\acrodef{DORA}{Digital Operational Resilience Act}
\acrodef{DSL}{Domain-Specific Language}
\acrodef{OECD}{Organisation for Economic Co-operation and Development}
\acrodef{EU}{European Union}
\acrodef{DG CONNECT}{Directorate-General for Communications Networks, Content and Technology}
\acrodef{ETSI}{European Telecommunications Standards Institute}
\acrodef{NLF}{New Legislative Framework}
\acrodef{MDR}{Medical Device Regulation}
\acrodef{RDF}{Resource Description Framework}
\acrodef{BNA}{Boundary Negotiating Artifact}
\acrodef{MLOps}{Machine Learning Operations}

\end{document}